\documentclass[revsymb,aps,prb,10pt]{revtex4}
\begin{document}
\draft
\title{Topological stripelike coreless textures with inner incommensurability in
two-dimensional Heisenberg antiferromagnet }
\author{E. V. Sinitsyn, I. G. Bostrem, A. S. Ovchinnikov\cite{byline}}
\address{Department of Physics, Ural State University, Lenin ave. 51, 620083\\
Ekaterinburg, Russia}
\date{\today }

\begin{abstract}
For two-dimensional Heisenberg antiferromagnet we present an
analysis of topological coreless excitations having a stripe form.
These textures are characterized by singularities at boundaries. A
detailed classification of the stripe textures results in a
certain analogy with the coreless excitations in $^3He-A\,$ phase:
Mermin-Ho and Anderson-Toulouse coreless vortices. The excitations
of the last type may have a low bulk energy. The stripe textures
may be observed as an occurrence of short-range incommensurate
order in the antiferromagnetic environment.
\end{abstract}

\maketitle

\section{Introduction}

The Heisenberg antiferromagnet (AFM) in two dimensions (2D) supports
nonlinear pseudoparticles with a vortex structure. An existence of these
excitations with a singular point, the center of the topological defect, is
issued from nontriviality of higher homotopic group $\pi _2(RP^2)=Z\,$ ($Z$
is an integer group) of the space $RP^2=O(3)/O(2)\times O(1)$ (or Grassmann
manifold $G_{3,1}$), the space of the antiferromagnetic order parameter $%
\vec{L}$. It is believed that these topological excitations are of
importance in the understanding of static and dynamical properties of 2D AFM.

The paper is devoted to another topological structures that may
occur in the system. Their appearance may be argued for the
following reasonings. The twofold degeneracy in the direction of
antiferromagnetic vector is a source for formation of a domain
structure with domain walls between N$\acute{e}$el-like ground
states. The domain walls have an energy scaling with a linear size
of the system. Apart from the domain structure formation there is
another reason for the field $\vec{L}$ to be inhomogeneous. One
may assume an appearance of a helical coreless spin texture
between two uniform N$\acute{e}$el-like ground states which is a
topological excitation, soliton, with finite energy. First, this
idea has been proposed in the study of spin properties of quantum
Hall (QH) states. A formation of low-energy topological
excitations localized in the domain walls between oppositely
polarized domains has been considered in the investigation of a
multidomain structure in a ferromagnetic QH liquid.\cite{Falko1}
Later, it has been shown that QH ferromagnets (QHF) with vanishing
Zeeman energy and a pronounced spin-orbit coupling are unstable
concerning to the formation of a helical state.\cite {Falko2}
Recently, a similar approach has been recurred for a QH Ising
ferromagnet at even filling factor.\cite{Brey} In the presence of
domains between ferromagnetic and unpolarized ground states,
charge excitations can be trapped in the walls forming confined
isospin textures, charged solitons at the domain wall. Due to
nonzero spin-orbit coupling the finite energy of such a soliton
has been found to be rather small.

The excitations with the same structure have been analyzed for a 2D Heisenberg
ferromagnet.\cite{Bostrem} Contrary to QHF they carry no electrical
charge and they are energetically expensive. It is very close to the
situation with Skyrmion-like textures. Whereas skyrmion/antiskyrmion pairs
may be thermally activated in QHF (Ref. 5) they actually freeze out in
ordinary 2D ferromagnets.

At last, we note that 2D stripe textures are well known in physics of
liquid crystals. Freely suspended smectic liquid-crystal films of HOBACPC
[R(-) hexyloxybenzylidene p'-amino-2-chloropropyl cinnamate] display
distinctive stripe textures. These patterns have been observed
experimentally.\cite{Langer}

Guided by the arguments we predict an appearance of stationary
stripelike coreless textures with an inner incommensurability in 2D
antiferromagnet. These textures may be viewed as excitations whose scale
along one direction in the plane coincides with the incommensurate
periodicity and, in other direction, they have soliton (kink) features. By
using a continuum approximation we classify these excitations and find
conditions needed for their appearance. We reveal that for the most
important types of the textures a space behavior of staggered magnetization
along one of the plane directions is akin to arrangement of order parameter
(angular moment of pair) in coreless vortices of $^3He-A$ phase, namely, in
Mermin-Ho (MH) \cite{MHO} and Anderson-Toulouse (AT) \cite{AT} vortices. Our
analysis shows that a stripe counterpart of AT vortex may have a low bulk
energy.

In contrary to excitations with a point singularity in a center an
appearance of the stripe excitations forms singular points at their
boundaries with a nonsingular (coreless) bulk structure inside. In the
theory of liquid helium this type of a "surface" singularity is known as
boojum.\cite{Mermin,Stein}

The paper is organized in the following way. In Sec. II the continuum
approximation based upon equations of nonlinear spin dynamics is presented.
In Secs. II.A-II.D the solutions with collinear antiferromagnetic,
spin-flop, spin-flip, and so-called "instanton-like" spin arrangements at
the texture outskirts are considered. Finally, in Sec. III we discuss a
possible application of the found textures for an explanation of
incommensurate (IC) correlations in cuprate materials of the spin glass (SG)
regime.\cite{Wakimoto,Wakimoto1,Matsuda}

\section{ Model}

The quantative analysis of a stripe texture is based on the Hamiltonian of a
spin-$S$ antiferromagnet

\begin{equation}
H=\frac 12J_{\perp }\sum\limits_{\left\langle m\alpha ,n\beta \right\rangle
}\left( S_{m\alpha }^xS_{n\beta }^x+S_{m\alpha }^yS_{n\beta }^y\right) +%
\frac 12J_z\sum\limits_{\left\langle m\alpha ,n\beta \right\rangle
}S_{m\alpha }^zS_{n\beta }^z-h\sum\limits_{m\alpha }S_{m\alpha }^z,
\label{Hamka}
\end{equation}
where $S_{m\alpha }^k$ is the $k$th component of the spin operator of the
$m$th site and $\alpha $ sublattice. The Hamiltonian has an exchange
anisotropy, the $J_{\perp }$, $J_z$ are the nearest neighbor exchange
integrals, and $\left\langle \ldots \right\rangle $ denotes the
sum over the nearest neighbor pairs. We have also included in Eq.(\ref{Hamka}%
) a Zeeman term with an external magnetic field $h$ along the $z$ axis.

By computing in spin-coherent representation an equation of motion for the
raising operator $i\hbar \left( dS_{m\alpha }^{\dagger }/dt\right) =\left[
S_{m\alpha }^{\dagger },H\right] $ $(\alpha =1,2)$ and going over to a
continuum approximation we get the coupled system of non-linear equations
for the variables $\theta _{1,2}$ and $\varphi _{1,2}$ that parametrize the
spin fields ${\vec S}_\alpha =S\left( \sin \theta _\alpha \cos \varphi
_\alpha ,\sin \theta _\alpha \sin \varphi _\alpha ,\cos \theta _\alpha
\right) $%
\[
\hbar \sin \theta _1\frac{d\varphi _{_1}}{dt}=-J_{\perp }S\left\{ \cos
\left( \varphi _2-\varphi _1\right) \cos \theta _1\sin \theta _2\left[
4-\left( \nabla \theta _2\right) ^2-\left( \nabla \varphi _2\right)
^2\right] \right.
\]
\[
+\cos \theta _1\cos \theta _2\cos \left( \varphi _2-\varphi _1\right) \Delta
\theta _2-\sin \left( \varphi _2-\varphi _1\right) \cos \theta _1\sin \theta
_2\Delta \varphi _2
\]
\[
\left. -2\sin \left( \varphi _2-\varphi _1\right) \cos \theta _1\cos \theta
_2\left( \nabla \varphi _2\nabla \theta _2\right) \right\}
\]
\begin{equation}
+J_zS\left\{ 4\sin \theta _1\cos \theta _2-\sin \theta _1\sin \theta
_2\Delta \theta _2-\sin \theta _1\cos \theta _2\left( \nabla \theta
_2\right) ^2\right\} -h\sin \theta _1,  \label{Eq1}
\end{equation}
\[
\hbar \frac{d\theta _1}{dt}=J_{\perp }S\left\{ \sin \left( \varphi
_2-\varphi _1\right) \sin \theta _2\left[ 4-\left( \nabla \theta _2\right)
^2-\left( \nabla \varphi _2\right) ^2\right] \right.
\]
\[
+\cos \left( \varphi _2-\varphi _1\right) \sin \theta _2\Delta \varphi
_2+\cos \theta _2\sin \left( \varphi _2-\varphi _1\right) \Delta \theta _2
\]
\begin{equation}
\left. +2\cos \left( \varphi _2-\varphi _1\right) \cos \theta _2\left(
\nabla \varphi _2\nabla \theta _2\right) \right\} ,  \label{Eq2}
\end{equation}
where a lattice constant is taken unit. The rearrangement of the lower
indices $1\rightleftharpoons 2$ yields another pair of equations.

We look for solitons having a stripelike texture. Below, we use the
parametrization $\theta _{1,2}=\theta _{1,2}(y)$ and $\varphi _{1,2}=\varphi
_{1,2}(x)$ with $\varphi _{1,2}\,$ obeying the constraint $\varphi
_2-\varphi _1=\pi $ over the 2D plane. That parametrization reduces Eq. (\ref
{Eq2}) to the simple equation $\triangle \varphi _{1,2}=0$. A suitable
solution may be taken in the form $\varphi _i=\varphi _{i0}+qx$ ($i=1,2$)
within an interval of length $2\pi /q$ ($0\leq x\leq 2\pi /q$) and as a
constant value $\varphi _{i0}$ out of that interval. The width of the stripe
texture is managed by a continuous parameter $q$. The $\theta _{1,2}(y)$
profiles in the stripe texture may be obtained from Eq. (\ref{Eq1})

\[
0=-h\sin \theta _2+J_{\perp }S\left\{ \cos \theta _2\sin \theta _1\left[
4-\left( \frac{d\theta _1}{dy}\right) ^2-q^2\right] +\cos \theta _2\cos
\theta _1\frac{d^2\theta _1}{dy^2}\right\}
\]
\begin{equation}
+J_zS\left\{ 4\sin \theta _2\cos \theta _1-\sin \theta _2\sin \theta _1\frac{%
d^2\theta _1}{dy^2}-\sin \theta _2\cos \theta _1\left( \frac{d\theta _1}{dy}%
\right) ^2\right\} ,\,(1\rightleftharpoons 2).  \label{tetastr}
\end{equation}
The degeneracy over the $q$ sign in Eq.(\ref{tetastr}) corresponds to a
clockwise or counter-clockwise spiral. Topological classes of spin textures with
the different chirality belong to the homotopic group $\pi _1(RP^2)=Z_2$ ($%
Z_2$ is a cyclic group).

At zero field a symmetry of the sublattices $\theta _2=\pi -\theta _1$
reduces the system (\ref{tetastr}) to the equation suitable for 2D
ferromagnet.\cite{Bostrem} At nonzero magnetic field one may expect a
small net magnetization, a weak ferromagnetism, due to a slight deviation of
the sublattice magnetization from antiparallel arrangement. Unfortunately,
an account of the magnetic field is essential and we have not been able to
obtain analytical solution. Therefore, we mention the results of numerical
investigation made by shooting method whenever it is necessary.

The first integral of the system (\ref{tetastr}) is readily derived. One
have to multiply the first equation by $d\theta _2/dy$, the second by $%
d\theta _1/dy$ and sum the results

\[
\left[ \cos \left( \theta _1+\theta _2\right) -\frac K{J_{\bot }}\sin \theta
_1\sin \theta _2\right] \frac{d\theta _1}{dy}\frac{d\theta _2}{dy}=
\]
\[
\left( \frac h{J_{\bot }S}\right) \left[ \cos \theta _{10}+\cos \theta
_{20}-\cos \theta _1-\cos \theta _2\right] +4\frac K{J_{\bot }}\left[ \cos
\theta _1\cos \theta _2-\cos \theta _{10}\cos \theta _{20}\right]
\]
\begin{equation}
+q^2\left[ \sin \theta _1\sin \theta _2-\sin \theta _{10}\sin \theta
_{20}\right] +4\left[ \cos \left( \theta _1+\theta _2\right) -\cos \left(
\theta _{10}+\theta _{20}\right) \right] .  \label{Int1}
\end{equation}
One can see this by nothing that a two-sublattice counterpart of Eqs. (4) and
(5.18) in Refs. 4, 14 , respectively. Hereinafter, the
anisotropy parameter $K=J_z-J_{\bot }$ denotes a difference between the
exchange integrals.

Our calculation of energetics of the topological textures is based on the
continuum approximation. For the bulk energy per stripe we find

\[
E=\int\limits_0^{2\pi /q}dx\int\limits_{-\infty }^\infty dy\;\omega
=\int\limits_0^{2\pi /q}dx\int\limits_{-\infty }^\infty dy\,\left\{ J_{\perp
}S^2\left[ \left( -4+q^2\right) \sin \theta _1\sin \theta _2+\cos \theta
_1\cos \theta _2\left( \nabla \theta _1\nabla \theta _2\right) \right]
\right.
\]
\[
+\left. J_zS\left[ 4\cos \theta _1\cos \theta _2-\sin \theta _1\sin \theta
_2\left( \nabla \theta _1\nabla \theta _2\right) \right] -hS\left( \cos
\theta _1+\cos \theta _2\right) \right\} .
\]
The soliton energy must be measured from the energy $E_0=\int\limits_0^{2\pi
/q}dx\int\limits_{-\infty }^\infty dy$ $\omega _0$ of a background spin
configuration corresponding to an arrangement at outskirts of the stripe.
Because of a twist in $\varphi _{1,2}$ angles the background spin order may
differ essentially from the ground state of the remaining system. The
background energy density is given by
\begin{equation}
\omega _0=J_{\perp }S^2\left( -4+q^2\right) \sin \theta _{10}\sin \theta
_{20}+4J_zS\cos \theta _{10}\cos _{}\theta _{20}-hS\left( \cos \theta
_{10}+\cos \theta _{20}\right) ,  \label{bgener}
\end{equation}
where $\theta _{10,20}$ are the constant values $\theta _{1,2}|_{y=\pm
\infty }$, respectively.

To evaluate the soliton energy density $\omega $ measured from $\omega _0$
it is convenient to use Eq.(\ref{Int1})
\[
\omega -\omega _0=2J_{\bot }S^2q^2\left( \sin \theta _1\sin \theta _2-\sin
\theta _{10}\sin \theta _{20}\right) +8J_{\bot }S^2\left( \cos \left( \theta
_1+\theta _2\right) -\cos \left( \theta _{10}+\theta _{20}\right) \right)
\]
\[
+8KS^2\left( \cos \theta _1\cos \theta _2-\cos \theta _{10}\cos \theta
_{20}\right) +2hS\left( \cos \theta _{10}+\cos \theta _{20}-\cos \theta
_1-\cos \theta _2\right) .
\]
In the expressions for the energy we neglect any surface terms.
Their appearance is associated with the abrupt $\theta _{1,2}$
behavior at the boundaries between the stripe texture and
surrounding N$\acute{e}$el-like ground states. A model estimation
of the surface contribution is given in Appendix A.

The stationary stripe textures are determined by the system (\ref{tetastr})
with the conditions $\theta _{1,2}|_{y=\pm \propto }=\theta _{10,20}$ and $%
(d\theta _{1,2}/dy)|_{y=\pm \infty }=0$%
\begin{equation}
\left\{
\begin{array}{c}
4\sin \left( \theta _{10}+\theta _{20}\right) -q^2\cos \theta _{10}\sin
\theta _{20}+4\frac K{J_{\bot }}\cos \theta _{20}\sin \theta _{10}-\frac h{%
J_{\bot }S}\sin \theta _{10}=0 \\
4\sin \left( \theta _{10}+\theta _{20}\right) -q^2\cos \theta _{20}\sin
\theta _{10}+4\frac K{J_{\bot }}\cos \theta _{10}\sin \theta _{20}-\frac h{%
J_{\bot }S}\sin \theta _{20}=0
\end{array}
\right. \text{,}  \label{boundsys}
\end{equation}
where $\theta _{10,20}$ are the boundary configurations. At $q=0$ each of
the configurations corresponds to a certain ground state of antiferromagnet
at nonzero magnetic field along $z$ axis.

The system (\ref{boundsys}) results in four types of boundary states. Here
we list these classes.

(I) $\theta _{10}=0$, $\theta _{20}=\pi $ (or vice versa). Collinear or
antiferromagnetic state with the opposite spins aligned along $z$ direction.

(II) $\theta _{10}=\theta _{20}=\theta _0$, $\cos \theta _0=h/\left(
8J_{\bot }S+4KS-J_{\bot }Sq^2\right) $. Spin-flop state with the canted
spins. The staggered magnetization is in the plane (basal plane)
perpendicular to a nonzero value of the total magnetization along $z$ axis.

(III) $\theta _{10}=0$, $\theta _{20}=0$ (or $\pi $). Spin-flip state with
the sublattice spins directed along $z$ axis.

(IV) $\theta _{10}\neq $ $\theta _{20}$. A class of instanton solutions with
the relationship
\begin{equation}
\theta _{10}=2\tan ^{-1}\left[ \cot \left( \frac{\theta _{20}}2\right) \frac{%
J_{\bot }S\left( 4-q^2\right) +4J_zS-h}{J_{\bot }S\left( 4-q^2\right)
+4J_zS+h}\right]  \label{angle12}
\end{equation}
and the restriction to the $q$ parameter
\[
q^2=4\pm \sqrt{16(J_z/J_{\bot })^2-\left( h/J_{\bot }S\right) ^2}.
\]

Depending on the value of the magnetic field different homogeneous states
outside of the stripe texture can be realized. We mention two characteristic
fields that may be obtained at $q=0$ from Eq. (\ref{bgener}). At zero field
in easy-axis regime ($K>0$) and in a magnetic field $h<h_{c1}=4S\sqrt{%
J_z^2-J_{\bot }^2}$ the uniform antiferromagnetic state has the lowest
energy. At the field $h_{c1}$ the antiferromagnetic vector "flops" down
onto the basal plane. In the region $h_{c1}<h<h_{c2}$ the total
magnetization increases with increasing field and finally at the
''exchange'' field $h_{c2}=4S(J_{\bot }+J_z)$ the spin-flop state
continuously transforms into the saturated spin-flip state with a maximal
total magnetization and zero staggered magnetization.

An asymptotic of Eq. (\ref{tetastr}) at $y\rightarrow \pm \infty \,$ allows
to build a phase diagram of the stripe textures. It depends on the
parameters $(J_{\bot },K,h)$ which span a three-dimensional phase space for
the solutions. Before giving the detailed analysis, let us point out general
features of possible spin textures in the system (see also Appendix B). In
the case of fixed field, the phase diagram in the plane $(q^2,K/J_{\bot })$
includes regions with different textures separated by "hypersurfaces" (Fig.\ref{diagram}).

In the small field limit $h<h_{c1}$ energetically unfavorable spin-flip
textures are supported over a wide range of twist parameter $q$ and
couplings $K/J_{\bot }$. The line with the parametric equation
\begin{equation}
q^2=4-\sqrt{16\left( 1+K/J_{\bot }\right) ^2-\left( h/J_{\bot }S\right) ^2}
\label{critlin}
\end{equation}
limits from below the region of collinear antiferromagnetic solitons. As one
can see the both textures can coexist in a vast area. There are also three
nonconnected regions of spin-flop excitations. The small $q$ region is of
special interest in view of continuum approximation validity and possible
applications of the theory to real systems. The instanton like solutions
with $\theta _{10}\neq $ $\theta _{20}$ realize at the critical line (\ref
{critlin}) dividing the spin-flop textures of small $q$ and the collinear
antiferromagnetic excitations.

By inspecting of the phase diagram in long-wave sector $q \to 0$ one may
find three distinct values of coupling constants:

\begin{itemize}
\item  {\ $K/J_{\bot }>-1+\sqrt{1+(h/4J_{\bot }S)^2}$ (easy-axis exchange)
corresponds to excitations with a collinear arrangement at the boundaries $%
y\to \pm \infty $, i.e. spin-flip and collinear antiferromagnetic
excitations; }

\item  {\ $-3/2+1/2\sqrt{1+(h/4J_{\bot }S)}\,<K/J_{\bot }<-1+\sqrt{%
1+(h/4J_{\bot }S)^2}$ (easy-plane exchange, predominantly) corresponds to
spin-flop excitations only; }

\item  {\ $-1<K/J_{\bot }<-3/2+1/2\sqrt{1+(h/4J_{\bot }S)}$. None of the
above stripe textures realize at small $q$, just short-range excitations
with essentially nonzero $q$ values are possible here. }
\end{itemize}

\subsection{Collinear antiferromagnetic texture}

The numerical integration of Eq.(\ref{tetastr}) obtained by shooting method
with the aid of linear approximation near zero point ($y=0$) $\theta
_1\approx c_1y$ and $\theta _2\approx \pi -c_2y$ (see also Appendix C)
yields the set of solutions $\theta _{1,2}(y)$ that may be classified as a
pair of kinks $(-\pi ,\pi )$ and $(2\pi ,0)$. These solutions have a range $%
2\pi $ over $y$ axis.

The in-plane magnetic arrangements of both sublattices are presented in Figs.%
\ref{AFM} (a,b). We note that the line $y=0$ may divide the regions with the
same or opposite in-plane spin directions along $y$ axis. We present here
the first case when one may use solutions $\theta _{1,2}$ in the unphysical
region [dotted lines in Fig \ref{AFM}(e)] together with the change $\varphi
_i\rightarrow \varphi _i+\pi $. This observation is based upon the trivial
relations $\sin \left( \pi -\delta \right) \cos \varphi _1=\sin \left( \pi
+\delta \right) \cos \left( \varphi _1+\pi \right) $, $\sin \left( \pi
-\delta \right) \sin \varphi _1=\sin \left( \pi +\delta \right) \sin \left(
\varphi _1+\pi \right) $, and $\cos \left( \pi -\delta \right) =\cos \left(
\pi +\delta \right) $.

At given $x$ coordinate the staggered magnetization vector does not incline
from a fixed angle with $y$ axis while the component $\vec{M}_{\bot }$ of
total magnetization changes its direction into opposite twice. The profile $%
L_z(y)$ [Fig.\ref{AFM}(f)] exhibits a Skyrmion-like behavior $L_z|_{y=0}=-1$
and $L_z|_{y=\pm \infty }=1$. The regions with a nonzero component $M_z$
occur as two symmetrical narrow bands around the line $y=0$. One can
understand this considering evolution of the relative spin orientation [Fig.%
\ref{AFM}(g)]. The state with a pure antiparallel orientation is broken by
an applied field. The unique coexistence of weak ferromagnetism and chiral
modulations in these solitons enables the occurrence of short-range
incommensurate structures with weak ferromagnetic moments.

This stripe texture is similar to Anderson-Toulouse coreless vortex texture
in superfluid $^3He-A$. Indeed, by moving in the real space along the $y$ axis
one maps the line into the path in the AFM order space which has the
topology of projective plane $RP^2$, the two-dimensional sphere $S^2$ with
identical diametrically opposite points on the surface. A path from $%
y=-\infty $ to the center $y=0$ is equivalent to the path between the two
identified poles on the sphere and topologically nontrivial, but a second
path from the center to the $y=+\infty $ returns the AFM\ vector to the
starting point and the resultant is equivalent topologically to no AFM
vector rotation at all. In other words, the Pontryagin topological index
\[
Q=\frac 1{8\pi }\int\limits_0^{2\pi /q}dx\int\limits_{-\infty }^{+\infty
}dy\,\varepsilon ^{\mu \nu }\,\vec{L}\cdot \left( \partial _\mu \vec{L}%
\right) \times \left( \partial _\nu \vec{L}\right) =0\text{.}
\]

For small fields $h<h_{c1}$ or, equivalently, for couplings
$K/J_{\bot }$ greater some critical value corresponding to
$h_{c1}$ the soliton bulk energy $E(q)$ has a minimum at small
$q$. At this point the energy gap between the collinear
N$\acute{e}$el-like ground state and the soliton has a minimal
value $E_{min}$. The gap value scales linearly with $q$ that
resembles a dependence on wave vector of ordinary spin-wave
Goldstone mode.

A phase transition outside of the stripe between the collinear
antiferromagnetic and the canted spin-flop ground states modifies the $q$ dependence of the soliton bulk energy in a drastic way (see Fig. 3). When a decreasing
of exchange coupling $K/J_{\bot }$ reaches a certain treshold of easy-axis
regime at given applied field or, equivalently, at $h>h_{c1}$ for a fixed $%
K/J_{\bot }$ the energy decreases gradually with a decreasing $q$. In the
lowest point $q_0=4-\sqrt{16\left( 1+K/J_{\bot }\right) ^2-\left( h/J_{\bot
}S\right) ^2}$ it equals zero. Nevertheless, the excitations occur to be
gapped. The point is that we define a soliton energy with regards to the
energy of stripe arrangement at $y=\pm \infty $, i.e. collinear Neel order
in this case. However, it is no  longer a ground state of the system.

\subsection{Spin-flop texture}

In this subsection we consider the spin-flop textures. In order to get $%
\theta _{1,2}(y)$ numerically we use the quadratic approximation in the
vicinity of zero point $\theta _1\approx \theta _0+c_1y^2$ and $\theta
_2\approx \theta _0-c_2y^2$. Obtained solutions have the form of kinks $%
(-\theta _0,2\pi -\theta _0)$ and $(2\pi -\theta _0,-\theta _0)$ of range $%
2\pi $ [Fig.\ref{Flop}(e)]. In order to keep the same directions of in-plane
sublattice magnetizations in the vicinity of the line $y=0$ it is convenient
to use unphysical values of $\theta _{1,2}(y)$ presented by dotted lines in
Fig \ref{Flop}(e). The movement into the unphysical region must be
simultaneous with the rotation of $\varphi _{1,2}$ by $\pi $.

At a given $x$ coordinate in-plane projections of staggered magnetization $\vec{L%
}_{\bot }$ $\,$point fixedly into one direction at any $y$ coordinate while $%
\vec{M}_{\bot }$ changes its direction at the line $y=0$ [Figs.\ref{Flop}%
(c-d)]. An evolution of the relative spin arrangement is depicted in Fig.\ref
{Flop}(g). The staggered magnetization of the initial configuration ($%
y=-\infty $) lies in the basal plane with a total magnetization $\vec{M}$
parallel to $z$ axis. A path from $y=-\infty $ to the center $y=0$ is
accompanied by a rotation of the sublattice spins. At first, the spins align
along $z$ axis then the initial configuration restores in the center and
further the rotation runs in reverse order. This explains the oscillating
behavior of $L_z$ component and $M_z$ profile which is entirely opposite to
the case of collinear antiferromagnetic texture [Fig.\ref{Flop}(f)].

To attain an analogy with vortex states in superfluid $^3He-A$ phase we note
a similarity between $\vec{L}(y)$ dependence and radial behavior of order
parameter in MH\ vortex in liquid helium. However, the spin-flop texture has
a more complex core structure where a canted spin arrangement repeats in the
center.

The $q$ dependences of the bulk energy per stripe are presented in Fig.\ref
{flopener}. The energy is measured from the background value which is a
twisted spin-flop phase with the period $q$. The energy turns into zero at
the line dividing the regions of spin-flop and collinear antiferromagnetic
excitations in the phase diagram.

\subsection{Spin-flip texture}

In this subsection we discuss properties of spin-flip textures. To obtain
the corresponding solutions of Eq.(\ref{tetastr}) one have to use the linear
approximations $\theta _1\approx \pi +c_1y$ and $\theta _2\approx \pi -c_2y$
near $y=0$. The $\theta _{1,2}(y)$ profiles may be classified as the $%
(0,2\pi )$ and $(2\pi ,0)$ kinks. The projections of the sublattice
magnetizations onto the plane are shown in Figs.\ref{Flip}(a-b). Contrary to
previous cases, the total magnetization $\vec{M}$\ becomes more essential in
comparison with the staggered magnetization $\vec{L}$. Both in-plane
arrangements of sublattice spins and the component $\vec{L}_{\bot }$ $\,$of
staggered magnetization have the same form as those for the collinear
antiferromagnetic excitations [Fig.\ref{Flip}(c)]. However, the $\vec{M}%
_{\bot }$ and $L_z$ components do not appear at all. The $M_z(y)$ dependence
exhibits a Skyrmion-like behavior: $M_z|_{y=0}=-1$ and $M_z|_{y=\pm \infty
}=1$ [Fig.\ref{Flip}(d)]. These features are easily explained by a
symmetrical deviation of the sublattice magnetizations from parallel
arrangement [Fig.\ref{Flip}e]. At last, we note that for small applied fields ( $h<h_{c1}$) an appearance
of the spin-flop texture leads to a great loss in energy of the system.

\subsection{Instanton like textures}

These solutions are similar to instanton-like excitations with a point
singularity for systems with an axial symmetry \cite{Bostrem1}. Indeed, one
of the variables, for example $\theta _{20}$, is determined according to Eq.(%
\ref{angle12}) by another variable $\theta _{10}$ which can be taken
arbitrarily in the range $(0,\pi )$. These solitons may exist only at line (%
\ref{critlin}) and require fulfillment of condition $h\leq 4J_zS$. An
instanton nature of these excitations lays in the fact that they provide a
gradual transition from collinear antiferromagnetic stripe textures, $(-\pi
,\pi )$ and $(2\pi ,0)$ kinks, to the spin-flop excitations, $(-\theta
_0,2\pi -\theta _0)$ and $(2\pi -\theta _0,-\theta _0)$ kinks. When $(-\pi
,\pi )$ kink shifts upward by an angle $\pi -\theta _0$ another $(2\pi ,0)$
kink displaces downward by an angle $\theta _0$ until the spin-flop texture
recovers.

\section{Conclusion}

Let us discuss briefly a possible application of the solutions found to real
systems. The physics of the intermediate doping regime of cuprate materials,
spin glass regime, is the hotly debated subject nowadays. The neutron
scattering data on $La_{2-x}Sr_xCuO_4$ have revealed incommensurate
correlations in this compound. The experiments have shown that the
correlation lengths in the SG\ regime ($0.02<x<0.05$) are extremely short
and of the same as the periodicity of the IC\ modulations. In addition,
there are experimental observation of macroscopic in-plane $a-b$ asymmetry
in transport and magnetic properties.\cite{Ando}

In early theory, Shraiman and Siggia predicted a forming of static spiral
spin correlations with a pitch proportional to the hole density but inside
the superconducting phase\cite{Schraiman}. For the insulating state, along
with the picture of well-ordered stripes, assuming an existence of a charge
order,\cite{Zaanen} a different explanation of the two IC peaks in the SG
phase has been proposed in the dipole model.\cite{Glazman,Aharony}
According to the model the IC signals may arise from the formation of a
spiral magnetic order which breaks $O(3)/O(2)$ symmetry of collinear AFM\
phase without invoking any kind of charge order.\cite{Neto} In the dipole
model the randomly distributed holes act as frustration centers for the
underlying antiferromagnetic background, generating dipole moment. A
fraction of these dipoles may order ferromagnetically, while the others may
remain disordered. To explain observable short range incommensurate
correlations it has been considered a phenomenological theory of
stabilization of incommensurate spiral configuration with nonzero average
twist of the AFM order and simultaneous alignment of the dipoles. In the
perturbative RG analysis the dipole disorder leads to a simple
renormalization of the spin stiffness which in its turn leads to a finite
correlation length already at $T=0$. However, a strongly disordered regime,
associated with a SG, is only found once topological defects of spin
textures are accounted for. In this approach an attention has been paid to
the topological defects analogous to that of the $XY$ model with the difference
that these topological defects have their origin in the chiral degeneracy of
the spiral ($Z_2$ defects).\cite{Kawamura}

In view of these investigations, it is not be excluded that the collinear
antiferromagnetic stripe texture may be relevant to physics of SG phase in
cuprates. These nonlinear excitations are essentially anisotropic, they have
a scale along the selected direction coinciding with a pitch of spiral, they
may possess an Izing-like anisotropy, they are stationary and has a bulk
energy smaller then topological structures with a core. A loss in the
surface energy provides a natural mechanism of a stripe attraction and leads
to an extension of region with the topological solitons. Due to the specific
structure (nonzero macroscopic ferromagnetic moment) one can admit that an
interaction between the spin texture and the external dipole subsystem
involves an additional dipole-dipole mechanism of energy decreasing, which
is of importance for undesirable surface energy. In the dipole model this
solves some principal difficulties with a simultaneous dipole ordering and a
fast correlation length growing in the spin spiral.

The collinear antiferromagnetic stripe texture has two boundary point
singularities at $x=0$ and $x=2\pi /q$ where the staggered magnetization
changes its direction. We call these points the surface $Z_2$ defects.
Then, one may say, the stripe texture occurs due to their presence. As for
ordinary Ising-like domain walls, an origin of the defects is caused by
nontriviality of the topological group $\pi _0(RP^2)=Z_2$ of the space $RP^2$
of AFM\ order parameter. We emphasize here a topological difference between
these $Z_2$ defects and those that have been used in the dipole model.\cite
{Neto} The latter describe a change of spiral chirality, clockwise or
counter-clockwise twist.

Finally, we discuss a dynamical stabilization of the stripe excitations. Let
us assume that stripe texture moves with a constant velocity $v$ along $x$
axis. We take into account the moving via the parametrization $\varphi
_1(x)=q(x-vt)$ and $\varphi _2(x)=q(x-vt)+\pi $. One may see that Eq.(\ref
{Eq1}) can be written as

\[
0=-\left( h-\hbar qv\right) \sin \theta _2+J_{\perp }S\left\{ \cos \theta
_2\sin \theta _1\left[ 4-\left( \frac{d\theta _1}{dy}\right) ^2-q^2\right]
+\cos \theta _2\cos \theta _1\frac{d^2\theta _1}{dy^2}\right\}
\]
\begin{equation}
+J_zS\left\{ 4\sin \theta _2\cos \theta _1-\sin \theta _2\sin \theta _1\frac{%
d^2\theta _1}{dy^2}-\sin \theta _2\cos \theta _1\left( \frac{d\theta _1}{dy}%
\right) ^2\right\} ,\,(1\rightleftharpoons 2).  \label{tetastr}
\end{equation}
i.e., the moving  is equivalent to an inclusion of an effective "magnetic
field" $-\hbar qv$ breaking the hidden chiral symmetry. Naturally, all
results found above may be extended for the case of mobile textures that may
occur at zero external magnetic field ($h=0$) due to the "self-focus"
effect. The conservation of macroscopic momentum of magnetization $P_x$
provides a mechanism of dynamical stability due to the relation
\begin{equation}
\delta E=\hbar Sqv\int dxdy\left( \sin \theta _1\delta \theta _1+\sin \theta
_2\delta \theta _2\right) =v\delta P_x  \label{DynStab}
\end{equation}
between variations of the energy $\delta E$ and the momentum $\delta P_x$
(see Appendix D).

Now, we consider briefly some experiments on doped cuprates in the spin-glass
regime. It is well known that the magnetism and the transport properties in
the doped $La_2CuO_4$ system are intimately related \cite{Kastner}.
Measurements of electrical resistivity in untwinned single crystals $%
La_{2-x}Sr_xCuO_4$ ($x=0.02-0.04$) give evidence that the doped electrons
self-organize into a macroscopically anisotropic state: the transport is
found to be easier along one of the plane direction, demonstrating that the
stripes are intrinsically conducting in cuprates.\cite{Ando} The resulting
in-plane anisotropy grows rapidly with decreasing temperature below $150 K$
and cannot be explained if one assumes that the anisotropy is due to simply
orthorombicity of the crystal. A mechanism responsible for the observed
anisotropy behavior is a remaining puzzle problem.

Extensive elastic neutron scattering studies have been performed on lightly
doped $La_{2-x}Sr_xCuO_4$ ($0\leq x\leq 0.055$) in order to elucidate the
static magnetic properties in the spin glass regime. The studies reveal that
the static spin correlations in the spin-glass phase show a one-dimensional
spin modulation. In the lightly doped regime $0\leq x\leq 0.02$, it is well
established that a three-dimensional (3D) antiferromagnetic long-range
ordered phase and a spin-glass phase coexist at low temperatures. Matsuda
{\it et al}. suggested that in this regime electronic phase separation of
the doped holes occurs so that some regions with hole concentration $c_h\sim
x$ exhibit incommensurate correlations while the rest with $c_h\sim 0$ shows
3D AF order \cite{Endoh}. Matsuda {\it et al}. also found that the magnetic
correlations change from being incommensurate to commensurate at $T\sim 70 K$
\cite{Matsuda}.

We suggest the following qualitative physical picture modeling a situation
in the lightly doped $La_2CuO_4$ system. Although the picture may not be the
only possible explanation for the specific behavior of the cuprates, the
following discussion shows that it does not contradict the experimental facts
mentioned above.

Let us assume that to gain in a kinetic energy the doped charge
may deforms homogeneous N$\acute{e}$el ground state into the
excited collinear antiferromagnetic stripe texture. The doped hole
can be trapped in the created stripe forming a charged soliton
akin to charged excitations in quantum Hall Ising ferromagnets. As
a hole travels it favors to pass across the stripe along regions
with nonzero ferromagnetic moments. During the process, the doped
hole may deforms adiabatically homogeneous Neel ground state
around that would be a source of a stripe movement. An effective
"magnetic field" $-\hbar qv$ originating from the nonzero stripe
velocity is responsible for an appearance of noncompensated
ferromagnetic moments inside the stripe. On the other hand, the
movement provides a mechanism of dynamical stabilization of the
new spin texture [Eq.(\ref{DynStab})]. The latter is directly
analogous to the scheme of "rotating bucket" used in experiments
with liquid helium.\cite{Leggett,Madison} Under this rotation, the
formation of vortices is, in principle, a consequence of thermal
equilibrium. Above a critical rotation frequency $\Omega _c$, the
term $-\Omega L_z$ in the Hamiltonian $\tilde{H}=H-\Omega L_z$,
where $H$ is the Hamiltonian in the absence of rotation, can favor
the creation of a state where the condensate wave function has an
angular momentum along $z$ axis and therefore contains a vortex
filament.\cite{Baym} The stripe spin texture moving with the
velocity $v$ provides a minimum of the functional
$\tilde{H}=H-vP_x$. Due to this reason, the collinear AFM texture
acts as a steady state supporting conserving momentum $\vec{P}$.
Then, it is natural to assume that the IC spin modulation observed
in elastic neutron scattering arises from the static correlations
of the steady stripe texture. On the contrary, only elementary
excitations, i.e. usual spin waves, carry nonzero momentum in the
case of uniform N$\acute{e}$el-like ground state and contribute to
inelastic magnetic spectra in neutron scattering measurements.

Ando {\it et al.} found that resistivity anisotropy $\rho _b/\rho _a$ in $%
La_{2-x}Sr_xCuO_4$ ($0.02\leq x\leq 0.04$) falls rapidly with
increasing temperature in the range $50\sim 100K$.\cite{Ando} We
may conclude that such a behavior strongly violates the dynamical
stabilization of the stripe texture due to decreasing of an
effective stripe velocity $v$. Without the moving, the stripe
texture is no longer a steady state and after the movement is
removed, one expects that the spins will eventually relax to
nontopological N$\acute{e}$el-like order. These arguments may
explain the results discovered by Matsuda {\it et
al.}\cite{Matsuda}{\it , }namely,{\it \ } why the magnetic
correlations change from being incommensurate to commensurate at
$\sim 70 K$ in $La_{2-x}Sr_xCuO_4$ ($0.02\leq x\leq 0.055$) .

A detailed quantative discussion of these effects is an important open
question and way beyond the scope of the present paper. We hope that this
discussion will stimulate some further studies of the stripelike nonlinear
textures and would be interesting for experimental researches.

In summary, an analysis of the stripe-like coreless textures in 2D
antiferromagnet is presented. The "kink" classification is given for these
nonlinear excitations. The topological mechanism explaining an appearance
of incommensurate quasi-one-dimensional structures in a 2D antiferromagnet is
suggested.

\acknowledgments
We would like to thanks V. Juricic, M. V. Sadovskii and M. V. Medvedev for the
useful discussion and interest in this study. This work was partly supported
by grant NREC-005 of US CRDF (Civilian Research \& Development
Foundation), INTAS grant (Project N 01-0654), by the grant "Russian
Universities" (UR.01.01.005).

\newpage {}

\appendix

\section{Surface energy}

Let us consider one stripe in the AFM\ environment. An additional surface
energy will be finite when a background state at the stripe outskirts ($%
y=\pm \infty $) coincides with the uniform state of the remaining AFM\
matrix. By assuming the ratio $\varphi _2-\varphi _1=\pi $ is kept over the
plane, i.e., inside and outside of the stripe, one obtain the linear density
of the surface energy
\[
J_{\bot }S_{2m}S_{1s}=J_{\bot }S^2\left[ \sin \theta _1\sin \theta _{20}\cos
\left( \varphi _{20}-\varphi _1\right) +\cos \theta _1\cos \theta
_{20}\right] =J_{\bot }S^2\cos \left( \theta _{20}+\theta _1\right) ,
\]
where the border spins $S_{2m}$ of the external AFM\ matrix interact with
the stripe border spins $S_{1s}$. For simplicity, we take an isotropic case.
If were the border line is being inside the stripe completely it would
contribute the value $J_{\bot }S^2\cos \left( \theta _2+\theta _1\right) $
into the energy. This expression accounts implicitly a similar spin
arrangement around the line. Hence, a contribution into the full energy,
that the surface brings in, has the form
\[
E_s=2J_{\bot }S^2\int\limits_{-\infty }^\infty dy\left[ \cos \left( \theta
_{20}+\theta _1\right) -\cos \left( \theta _2+\theta _1\right) \right] ,
\]
where $E_s$ is twice as much of the result for one border. The full energy
associated with the stripe is the sum of the bulk and the surface energies.
Now, we estimate $E_s$ for the collinear antiferromagnetic\ solitons with $%
\theta _{20}=\pi $ by assuming a step-like approximation, $\theta _1(y)=\pi $
at $\left| y\right| \leq \lambda ^{-1}$ and $\theta _1(y)=0$ at $\left|
y\right| >\lambda ^{-1}$, that yields immediately the rude estimation $%
E_s=16\pi J_{\bot }S^2/\lambda $.

\section{Phase diagram}

In order to get an asymptotic expansion at large distances we suppose in Eq.(%
\ref{tetastr}) $\theta _1=\theta _{10}+\delta _1$ and $\theta _2=\theta
_{20}+\delta _2$, where $\delta _1=A\exp (-\lambda y)$ and $\delta _1=B\exp
(-\lambda y)$ are small additions to the boundary values $\theta
_{10},\theta _{20}$. The linearization of the system gives
\[
\delta _1\left[ -(4-q^2)\sin \theta _{10}\sin \theta _{20}+4J_z/J_{\bot
}\cos \theta _{10}\cos \theta _{20}-h/J_{\bot }S\cos \theta _{10}\right]
\]
\[
+\delta _2\left[ (4-q^2)\cos \theta _{10}\cos \theta _{20}-4J_z/J_{\bot
}\sin \theta _{10}\sin \theta _{20}\right] +\left[ \cos \theta _{10}\cos
\theta _{20}-J_z/J_{\bot }\sin \theta _{10}\sin \theta _{20}\right] \frac{%
d^2\delta _2}{dy^2}=0,
\]

\[
\delta _1\left[ (4-q^2)\cos \theta _{10}\cos \theta _{20}-4J_z/J_{\bot }\sin
\theta _{10}\sin \theta _{20}\right]
\]
\[
+\delta _2\left[ -(4-q^2)\sin \theta _{10}\sin \theta _{20}+4J_z/J_{\bot
}\cos \theta _{10}\cos \theta _{20}-h/J_{\bot }S\cos \theta _{20}\right]
\]
\[
+\left[ \cos \theta _{10}\cos \theta _{20}-J_z/J_{\bot }\sin \theta
_{10}\sin \theta _{20}\right] \frac{d^2\delta _1}{dy^2}=0.
\]
The requirement of exponential decay at $y\rightarrow \infty $ yields a
parametric region where excitations with given $\theta _{10,20}$ exist. The $%
\lambda $ coefficient may be found from
\[
\left[ (4+\lambda ^2-q^2)\cos \theta _{10}\cos \theta _{20}-(4+\lambda
^2)J_z/J_{\bot }\sin \theta _{10}\sin \theta _{20}\right] ^2=
\]
\begin{eqnarray*}
&&\left[ (q^2-4)\sin \theta _{10}\sin \theta _{20}+4J_z/J_{\bot }\cos \theta
_{10}\cos \theta _{20}-h/J_{\bot }S\cos \theta _{10}\right] \times \\
&&\left[ (q^2-4)\sin \theta _{10}\sin \theta _{20}+4J_z/J_{\bot }\cos \theta
_{10}\cos \theta _{20}-h/J_{\bot }S\cos \theta _{20}\right] .
\end{eqnarray*}
The relation between $A$ and $B$ is given by
\[
\left[ -(4-q^2)\sin \theta _{10}\sin \theta _{20}+4J_z/J_{\bot }\cos \theta
_{10}\cos \theta _{20}-h/J_{\bot }S\cos \theta _{10}\right] A+
\]
\[
\left[ (4+\lambda ^2-q^2)\cos \theta _{10}\cos \theta _{20}-(4+\lambda
^2)J_z/J_{\bot }\sin \theta _{10}\sin \theta _{20}\right] B=0.
\]
One have to choose between two variants of $\lambda ^2$ those that provides
a different sign of $A$ and $B$ amplitudes.

For the collinear antiferromagnetic\ texture
\[
\lambda ^2=q^2-4+4\sqrt{\left( 1+K/J_{\bot }\right) ^2-\left( h/4J_{\bot
}S\right) ^2}.
\]
The corresponding region with $\lambda ^2\geq 0$ lies above the hyperbola
\[
\frac{\left( 1+K/J_{\bot }\right) ^2}{\left( h/4J_{\bot }S\right) ^2}-\frac{%
\left( q^2-4\right) ^2}{\left( h/J_{\bot }S\right) ^2}=1
\]
in the phase diagram. The domain is dashed horizontally in Fig.\ref{diagram}.

The flop-type excitations ($\theta _{10}=$ $\theta _{20}=\theta _0$) with
\[
\lambda ^2=\frac{q^2+4K/J_{\bot }-\left( h/J_{\bot }S\right) \cos \theta _0}{%
\cos ^2\theta _0-\left( 1+K/J_{\bot }\right) \sin ^2\theta _0}
\]
are supported by three regions determined by the conditions $\lambda ^2\geq
0 $ and $\left| \cos \theta _0\right| \leq 1$. The relevant domain is shown
in the phase diagram by the grey filling. The region of the spin-flip
textures with $\lambda ^2=q^2+4K/J_{\bot }-h/\left( J_{\bot }S\right) $ lies
over the line $\lambda ^2=0$. It should be noted here an existence of
regions where solitons of several types exist simultaneously.

\section{Approximation of starting values.}

In a numerical calculation by shooting method one have to use suitable
series expansions for $\theta _{1,2}\,$variables in the vicinity of the line
$y=0$
\[
\theta _i\approx \theta _{i0}+c_{i1}y+c_{i2}y^2+c_{i3}y^3\;(i=1,2).
\]
By substituting these expressions into the system (\ref{tetastr}) one may
obtain relations between the different coefficients and determine which of
them equal to zero. This procedure gives the following result for the AFM\
texture ($\theta _{10}=0$, $\theta _{20}=\pi $)
\[
-2J_{\bot }Sc_{22}+\left\{ -c_{11}h+J_zS(-4c_{11}+c_{11}c_{21}^2)+J_{\bot
}S\left[ -6c_{23}-c_{21}(4-c_{21}^2-q^2)\right] \right\} y+O(y^2)=0,
\]
\[
-2J_{\bot }Sc_{12}+\left\{ c_{21}h+J_zS(-4c_{21}+c_{21}c_{11}^2)+J_{\bot
}S\left[ -6c_{13}-c_{11}(4-c_{11}^2-q^2)\right] \right\} y+O(y^2)=0
\]
that provides $c_{12}=c_{22}=0\,$ and we may restrict by the linear
approximation $\theta _i\approx \theta _{i0}+c_{i1}y$ in a numerical study.
The same reasonings are suitable for the spin-flip excitations ($\theta
_{10}=\theta _{20}=\pi $) that yields $\theta _i\approx \pi +c_{i1}y$.
However, for the spin-flop excitations ($\theta _{10}=\theta _{20}=\theta _0$%
) we obtain the relations
\[
c_{i1}^2=4\frac{J_{\bot }-J_z}{J_{\bot }+J_z}\cot \left( 2\theta _0\right)
\,c_{i2}.
\]
At given second coefficient $c_{i2}$ the first coefficient $c_{i1}$ is
negligible for the small exchange anisotropy $K=J_{\bot }-J_z$ and far from
the values $\theta _0=0$ and $\pi $. This observation allows one to use a
quadratic approximation $\theta _i\approx \theta _0+c_{i1}y^2$ for numerical
studies.

\section{The momentum of magnetization in the stripe.}

The Lagrangian density of the system
\[
L=\sum\limits_{i=1}^2\hbar S\left( \cos \theta _i-1\right) \frac{\partial
\varphi _i}{\partial t}-J_{\bot }S^2\left[ 4\sin \theta _1\sin \theta _2\cos
\left( \varphi _1-\varphi _2\right) +\sin \theta _1\cos \theta _2\sin \left(
\varphi _1-\varphi _2\right) \left( \nabla \theta _2\nabla \varphi _1\right)
\right.
\]
\[
-\sin \theta _1\sin \theta _2\cos \left( \varphi _1-\varphi _2\right) \left(
\nabla \varphi _1\nabla \varphi _2\right) +\cos \theta _1\sin \theta _2\sin
\left( \varphi _2-\varphi _1\right) \left( \nabla \theta _1\nabla \varphi
_2\right)
\]
\[
\left. -\cos \theta _1\cos \theta _2\cos \left( \varphi _2-\varphi _1\right)
\left( \nabla \theta _1\nabla \theta _2\right) \right]
\]
\[
-J_zS^2\left[ 4\cos \theta _1\cos \theta _2-\sin \theta _1\sin \theta
_2\left( \nabla \theta _1\nabla \theta _2\right) \right] +hS\left( \cos
\theta _1+\cos \theta _2\right)
\]
allows us to get a momentum
\[
P_k=\hbar S\sum\limits_{i=1}^2\left( 1-\cos \theta _i\right) \frac{\partial
\varphi _i}{\partial x_k},\;(k=x,y)
\]
by using the Noether operator
\[
N^t=-\sum\limits_{i=1}^2\theta _{ik}\frac \partial {\partial \theta _{it}}%
-\sum\limits_{i=1}^2\varphi _{ik}\frac \partial {\partial \varphi _{it}}
\]
on the density $L$.\cite{Egorov}

The full macroscopic momentum for the stripe texture with $\varphi
_{ik}=q\delta _{kx}\,$%
\[
P_k=\delta _{kx}\;\hbar Sq\sum\limits_{i=1}^2\left( 1-\cos \theta _i\right)
.
\]
One has to measure the momentum from the macroscopic momentum $P_k^0$ of a
background order realizing at the stripe outskirts. The background momentum
of collinear antiferromagnetic texture is $P_k^0=2\hbar Sq\delta _{kx}$ that
results in the relative momentum
\[
\triangle P_k=-\hbar Sq\left( \cos \theta _1+\cos \theta _2\right) \delta
_{kx}
\]
as associated with the stripe.

\newpage {}

\begin{figure}[tbp]
\caption{The phase diagram of the stripe textures. Magnetic field $%
h/(4J_{\bot }S)=0.1$.}
\label{diagram}
\end{figure}

\begin{figure}[tbp]
\caption{Collinear antiferromagnetic stripe texture: in-plane arrangement of
sublattice magnetizations (a,b), total magnetization (c) and staggered
magnnetization (d). The $\theta _{1,2}(y)$ profiles (e), $L_z$ and $M_z$
components (f). An evolution of relative spin arrangement along the y axis (g).}
\label{AFM}
\end{figure}

\begin{figure}[tbp]
\caption{The $q$ dependence of bulk energy for the collinear
antiferromagnetic stripe texture at $h/(4J_{\bot }S)=0.1$ (a); the $q$ %
dependence of the energy gap $E_{min}$ (b), numbers in the plot point the $%
K/J_{\bot }$ ratio.}
\label{AFMener}
\end{figure}

\begin{figure}[tbp]
\caption{Spin-flop stripe texture: in-plane arrangement of sublattice
magnetizations (a,b), total magnetization (c) and staggered magnetization
(d). The $\theta _{1,2}(y)$ profiles (e) and the components $L_z$ and $M_z$
(f). An evolution of relative spin arrangement along the y axis (g).}
\label{Flop}
\end{figure}

\begin{figure}[tbp]
\caption{The $q$-dependence of bulk energy for the spin-flop stripe texture
at $h/(4J_{\bot }S)=0.1.$}
\label{flopener}
\end{figure}

\begin{figure}[tbp]
\caption{Spin-flip stripe texture: in-plane arrangement of sublattice
magnetizations (a,b), staggered magnetization vector $L_{\bot }$ (c) and $%
M_z(y)$ profile (d). A relative spin arragement along the y axis is shown below
(e).}
\label{Flip}
\end{figure}

\end{document}